\documentclass[aip,jcp,preprint,noshowkeys,superscriptaddress]{revtex4-1}
 
\usepackage{graphicx,bm,xcolor,microtype,multirow,amscd,amsmath,amssymb,amsfonts,physics,longtable,wrapfig,txfonts,soul}

\usepackage[utf8]{inputenc}
\usepackage[T1]{fontenc}
\usepackage{txfonts}

\usepackage[
	colorlinks=true,
    citecolor=blue,
    breaklinks=true
	]{hyperref}
\DeclareMathOperator*{\argmin}{argmin} 

\newcommand{\evGW}[0]{\mathrm{ev}GW}
\newcommand{\qsGW}[0]{\mathrm{qs}GW}
\newcommand{\scGW}[0]{\mathrm{sc}GW}
\newcommand{\qsGWJAD}[0]{\mathrm{qs}GW_\mathrm{JAD}}
\newcommand{\gsGWJAD}[0]{\gamma\mathrm{s}GW_\mathrm{JAD}}

\begin{document}	

\title{ Joint Approximate Diagonalization  approach to Quasiparticle  Self-Consistent $GW$ calculations }

\author{Ivan Duchemin}
\affiliation{Univ. Grenoble Alpes, CEA, IRIG-MEM-L\_Sim, 38054  Grenoble, France}
\email{ivan.duchemin@cea.fr}

\author{Xavier Blase}
\affiliation{ Univ. Grenoble Alpes, CNRS, Institut N\'{e}el, F-38042 Grenoble, France }

\date{\today}

\begin{abstract}
 We introduce an alternative route to quasiparticle self-consistent $GW$ calculations ($\qsGW$) on the basis of a Joint Approximate Diagonalization of the one-body $GW$ Green's functions $G(\varepsilon_n^{QP})$ taken at the input quasiparticle energies. Such an approach allows working with the full dynamical self-energy,   without approximating the latter   by a symmetrized static form as in the standard $\qsGW$ scheme. Calculations on the $GW$100 molecular test set lead nevertheless to a good  agreement,  at the 65 meV mean-absolute-error accuracy on the ionization potential,  with respect to the conventional  $\qsGW$ approach. We show further that constructing the density matrix from the full Green's function as in the fully self-consistent  $\scGW$ scheme,  and not from the occupied quasiparticle one-body orbitals,   allows   obtaining a scheme intermediate between $\qsGW$ and $\scGW$ approaches,  closer to CCSD(T) reference values.  
\end{abstract}


\maketitle
 
\section{ Introduction }

The Green's function many-body $GW$ perturbation theory \cite{Hed65,Gol19rev,Bruneval2021,ReiningBook} has become a tool of choice in condensed-matter physics for the description of charged excitations,  i.e. the electronic energy levels as obtained with photoemission experiments. Following pioneering application to the electron gas \cite{Hed65} or simple semiconductors, insulators and polymers, \cite{Str80,Hyb86,God88,Lin88} the $GW$ formalism is now being used with much success as well in the study of molecular systems where it can be compared to other perturbative approaches such as coupled-cluster techniques. \cite{Lange2018,Quintero2022,Tolle2023}  The favorable scaling of the $GW$ formalism, from quartic in its most common resolution-of-the-identity implementation, \cite{Ren2012} to cubic and below, adopting space-time, \cite{Rojas1995,Foerster2011,Kutepov2012,Wilhelm2018,Kim2020,Forster2020,Duchemin2021}  moment-conserving \cite{Scott2023}  or stochastic \cite{Neuhauser2014,Vlcek2017} techniques, contributes significantly to the success of this approach, together with the ability to deal with finite size or periodic, insulating or metallic, systems.    

In its most common historical implementation, the needed time-ordered Green's function $G$ and independent-electron susceptibility $\chi_0$, used to build the screened Coulomb potential $W$, are constructed from input one-body orbitals generated within mean-field Hartree-Fock (HF) or Kohn-Sham density functional theory (DFT). Such a relatively simple scheme is labeled the non-self-consistent, or single-shot, $G_0W_0$ formalism, an efficient approach that however provides results that depend on the input Kohn-Sham orbitals. The $G_0W_0$ results may prove not as reliable as needed if the input mean-field solution (charge-density, energy levels, orbitals shape, etc.) turns out to be very inaccurate. 

A first strategy to improve the accuracy of $G_0W_0$ calculations consists in optimizing the input mean-field orbitals using, e.g., optimally tuned global or range-separated hybrids. 
The amount of exact exchange or the range-separation parameter can be tuned by satisfying e.g. the condition that the negative of the Kohn-Sham highest occupied orbital energy matches the ionization potential calculated within a more accurate $\Delta$SCF calculation.\cite{Stein2010} Such an approach is very efficient and popular in particular in the case of molecular systems. \cite{Blase2011a,Korzdorfer2012,Bruneval2013,Rangel2016,Kaplan2016} 

In the case of infinite periodic systems, the difficulties associated with $\Delta$SCF calculations, and the cost of calculating the exact exchange contribution  as compared to purely (semi)local functionals, lead to consider instead self-consistent schemes.  Very early, \cite{Hyb86}  a partially self-consistent scheme, with update of the quasiparticle energies only,  was suggested, and studied extensively for solids \cite{Surh1988,Rohlfing1997,Shishkin2007} and molecular systems. \cite{Rangel2016,Kaplan2016} Such a simple self-consistent scheme is often labeled  $\evGW$ and can be further simplified in a scissor-like approach. \cite{Baumeier2012,Filip2014,Vlcek2018}  

Full self-consistency beyond the $\evGW$ scheme  was first introduced for a simple 1D model system, \cite{deGroot1995}   the interacting homogeneous electron gas at varying density  \cite{Holm1998,Gonzalez2001} and simple metals or semiconductors. \cite{Schone1998,Ku2002,Chu2016,Grumet2018,Kutepov2022,Yeh2022} In such studies, the Green's function is updated self-consistently together with the screened Coulomb potential treated as a functional of $G$ through the dependence of the independent electron susceptibility on $G$
(in short, $i\chi_0 = GG$). The self-consistent variable is thus the non-local energy-dependent $G({\bf r},{\bf r}'; E)$ time-ordered Green's function, bypassing the need for  one-body orbitals. This defines the fully self-consistent $\scGW$ approach that was used extensively to study molecular systems. \cite{Thygesen2008,Stan2009,Rostgaard2010,Marom2012,Caruso2012,Caruso2013,Koval2014,Wang2015,Knight2016,Caruso2016,Hung2016,Wen2024,Zgid2024} 
Extensions to relativistic $\scGW$ schemes were further proposed for molecules containing heavy elements. \cite{Abraham2024}

The fully self-consistent $\scGW$ was dramatically simplified by Faleev and coworkers who introduced a constrained self-consistency preserving the use of one-body wavefunctions. \cite{Faleev2004,Schilfgaarde2006} Such a quasiparticle self-consistent ($\qsGW$) approach relies on an \textit{ansatz} energy-independent symmetrized self-energy that upon diagonalization provides updated quasiparticle one-body wavefunctions and energies. This approach generalizes an early strategy where self-consistency was performed with the static Coulomb-Hole plus Screened Exchange (COHSEX) limit to the self-energy. \cite{Bruneval2006} An equivalent Hamiltonian could be obtained by minimizing the total energy expressed as a functional of the one-body Green's function, adopting the so-called Klein functional and minimizing over non-interacting Green's functions. \cite{Ismail-Beigi_2017}
The $\qsGW$ scheme was extensively applied to solids \cite{Faleev2004,Schilfgaarde2006,Svane2010,Punya2012,Bruneval2014,Brivio2014,Grumet2018,Lei2022,Zgid2024}  and molecules, \cite{Ke2011,Koval2014,Caruso2016,Kaplan2016,Forster2021b,Forster2022,Zgid2024} with extensions to two-components 2C-$\qsGW$  for molecules with heavy elements. \cite{Forster2023} A different $\qsGW$ scheme was introduced by Kutepov \textit{et al.} \cite{Kutepov2012} leading to a static self-energy through a linearization of its fully dynamical expression.   For molecular systems, this second $\qsGW$ scheme was found to yield results close to the  Shilfgaarde-Kotani-Faleev $\qsGW$ scheme. \cite{Zgid2024} 

Alternatively,  Loos and coworkers introduced a similarity renormalization group approach to Green's function methods, \cite{Monino2022,Marie_2023} leading to a regularized definition of a static and hermitian self-energy, allowing to set up another formulation of a quasiparticle self-consistent $GW$ scheme. Such an approach was further shown to cure difficulties associated with $GW$ spectral functions dominated by several peaks. \cite{Veril2018,Monino2022,Marie_2023}

A significant observation is that the $\scGW$ and $\qsGW$ ionization potential (IP) of small molecules was shown to differ by several tenths of an eV, \cite{Koval2014,Caruso2016} with e.g. an average difference of 0.59 eV for the five primary nucleobases as part of the $GW$100 test set. \cite{vanSetten2015,Krause2015,Caruso2016,Maggio2017,Govoni2018,Gao2019,Forster2021a} Overall, for the $GW$100 test set, the IP mean-signed-error (MSE) was found to amount to -0.30/+0.15~eV for $\scGW$/$\qsGW$ as compared to CCSD(T) calculations. \cite{Caruso2016} In a recent study performed on a partially different set of 29 small molecules, \cite{Maggio2017} the $\scGW$ scheme was found slightly more accurate than the $\qsGW$ scheme, with MAE on IPs of 0.24 and 0.28-0.29 eV, respectively, as compared to CCSD(T) calculations. \cite{Zgid2024} In agreement with the $GW$100 case, the $\scGW$/$\qsGW$ schemes were found to underestimate/overestimate the IP values as compared to CCSD(T). The $\scGW$ IPs of a set of larger acceptor molecules was also found to be systematically underestimated as compared to CCSD(T) reference calculations with a mean absolute error (MAE) of 0.6~eV, while on the contrary the electronic affinity (AE) was overestimated with a similar 0.61 eV MAE. \cite{Knight2016} The origin of these differences between the two self-consistent approaches was tentatively analyzed in terms of screening, charge density differences and different treatment of the kinetic energy. \cite{Caruso2016}

We present in this study an alternative quasiparticle self-consistent $GW$ scheme that does not rely on a symmetrized static self-energy \textit{ansatz} but introduces the idea of Joint Approximate Diagonalization \cite{Cardoso1996} (JAD) of the $G(\varepsilon_n^{QP})$ set of Green's functions, with   $\lbrace \varepsilon_n^{QP} \rbrace$ the quasiparticle energies given as input. Such an approach  provides the optimal one-body molecular orbitals  maximizing  the diagonality of the Green's function taken at the quasiparticle energies. As a central issue, the present scheme allows working with the full dynamical self-energy without any simplification.  Our $\qsGWJAD$ scheme is benchmarked over the $GW$100 molecular test set for which reference def2-TZVPP CCSD(T), \cite{Krause2015} $\qsGW$ and $\scGW$ calculations \cite{Caruso2016} are available. Remarkably, even though relying \textit{a  priori} on a very distinct  rational, the present quasiparticle self-consistent approach yields  ionization potentials in good agreement with the conventional $\qsGW$ quasiparticle self-consistent approach. Further, updating the  density matrix through integrating the Green's function along the imaginary axis, rather than  summing the contributions from the occupied one-body orbitals, in a simple scheme intermediate between $\qsGW$ and $\scGW$, is shown to yield a better agreement with CCSD(T) data, reducing the mean-signed-error to about 60 meV for the $GW$100 test set ionization potentials.  

\section{Theory}

In this Section, we outline the very basics of the $GW$ formalism,    \cite{Ary98,Farid99,Oni02,ReiningBook,Pin13,Gol19rev} 
and start by discussing the traditional  diagonal approximation for the self-energy expressed in the Kohn-Sham molecular orbitals basis. We then present non-self-consistent and self-consistent schemes where the quasiparticle energies are extracted from the Green's function spectral function, rather than from the quasiparticle equation, without the need to assume any restricted form (diagonal, static and symmetrized, etc.) for the self-energy.

\subsection{ The $GW$  formalism }

Belonging to the family of Green's function many-body perturbation theories, the $GW$ formalism takes as a central variable the one-body time-ordered Green's function: 
 \begin{align}
 G({\bf r},{\bf r}'; \omega ) =  \sum_{n}    \frac{   g_n({\bf r})  g_n^*( {\bf r}')  }{ {\omega} - \varepsilon_n  +i \eta \times \text{sgn}(\varepsilon_n  -E_F) }  
 \label{eqn:GreenFunc}
 \end{align}
 with $\eta$ a positive infinitesimal and $E_F$ the Fermi energy. Formally, the functions $\lbrace g_n({\bf r}) \rbrace$ are Lehman weights that measure how the ground-state Fermi sea with one added electron/hole  in  (${\bf r}$) overlaps with the (N+1)/(N-1)-electrons n-th excited state. Similarly, the energy poles $\lbrace \varepsilon_n   \rbrace$ are the proper charging energies, as measured e.g. by photoemission, namely the difference of total energy $E_n(N+1)-E_0(N)$   for empty states, and $E_0(N)-E_n(N-1)$ for occupied levels. Even in a finite size basis set of dimensionality $N_{basis}$, the number of poles  of $G(\omega)$  can be larger than $N_{basis}$. Anticipating on the $GW$ approximation to the self-energy, it was shown that the number of poles formally scales as the size of the one-hole-two-electron (2e-h) or two-electron-one-hole (e-2h) spaces. \cite{Backhouse2020,Bintrim2021,Monino2022}  

Following Hedin, \cite{Hed65} a set of self-consistent equations can be formally derived, relating  the electronic susceptibility operator $\chi$, the screened Coulomb potential $W$, the Green's function $G$,   the exchange-correlation self-energy operator $\Sigma$, and the higher-order (3-body) vertex correction $\Gamma$. To lowest order in the screened Coulomb potential, neglecting vertex corrections,   the energy-dependent exchange-correlation self-energy $\Sigma({\bf r},{\bf r}' ; E)$ can be written under the form of the $GW$ approximation : 
 \begin{align}
\Sigma^{GW}({\bf r},{\bf r}' ; E) = \frac{i}{2 \pi} \int_{-\infty}^{+\infty} d\omega \;  e^{i  \eta \omega  } G({\bf r},{\bf r}' ; E+ \omega) {W}({\bf r},{\bf r}' ;  \omega)
\label{eq:sigmaGW}
\end{align}
 with  $W$ the dynamically screened Coulomb potential built within the (direct) random phase approximation (RPA) : 
 \begin{align}
   {W}({\bf r},{\bf r}' ;  \omega) & = {V}({\bf r},{\bf r}' ) \nonumber \\
                 & + \int d{\bf r}_1 d{\bf r}_2   {V}({\bf r},{\bf r}_1 )   \chi_0({\bf r}_1 ,{\bf r}_2 ;  \omega)  \; {W}({\bf r}_2,{\bf r}' ;  \omega)  \label{eqn:Dyson} 
 \end{align}
 with $\chi_0({\bf r}_1 ,{\bf r}_2 ;  \omega)$ the independent-electron susceptibility and $V$ the bare Coulomb potential. The $GW$ approximation with the RPA screened Coulomb potential only contains ring diagrams and comparisons with other Green's functions approaches and coupled-cluster techniques have been proposed.   \cite{McClain2016,Lange2018,Quintero2022,Monino2023}
 
 In practice, the needed input Green's function and independent-electron susceptibility required to build a first approximation to the $GW$ self-energy are constructed with input Kohn-Sham eigenstates $\lbrace \varepsilon_n^{KS} , \phi_n^{KS}  \rbrace$, with e.g.:
 \begin{equation}
     \chi_0({\bf r},{\bf r}' ; \omega)  
    =   \sum_{mn}  \frac{ (f_m - f_n)    \phi_m^{KS}({\bf r})^{*}   \phi_n^{KS}({\bf r})        \phi_m^{KS}({\bf r}')  \phi_n^{KS}({\bf r}')^{*}   }{ \omega - ( \varepsilon_n^{KS} - \varepsilon_m^{KS}) + i\eta \times \text{sgn}{( \varepsilon_n^{KS} - \varepsilon_m^{KS} )} }  
    \label{eqn:xi0}
\end{equation}
 with $\lbrace f_{m/n} \rbrace$ level occupation numbers. Similarly, an input Green's function can be obtained by replacing the Lehman weights and charging energies in equation~\ref{eqn:GreenFunc} by the Kohn-Sham molecular orbitals and electronic energy levels. This construction leads to a starting $G_0W_0$ ``single-shot" self-energy that may strongly depend on the choice of the starting functional used to generate the input Kohn-Sham eigenstates. This dependence may be cured by performing self-consistent $GW$ calculations as discussed below.
 
 \subsection{ The diagonal self-energy approximation }
 
Stemming from historical calculations in simple systems such as bulk silicon, \cite{Hyb86,God88} it is commonly accepted  that, in general, the Hamiltonian built with the $GW$ self-energy is dominantly diagonal in the input Kohn-Sham basis. As such, the most common $GW$ calculations adopt a diagonal approximation where the contribution from the spurious DFT exchange-correlation potential is replaced by the expectation value of the $GW$ self-energy on the corresponding input Kohn-Sham eigenstate:
 \begin{align}
  \varepsilon_n^{GW} = \varepsilon_n^{KS} +  \langle \phi_n |  \Sigma^{GW}_{XC}(  \varepsilon_n^{GW} ) - V^{DFT}_{XC} | \phi_n \rangle
  \label{eqn:qpequation}
\end{align}
No off-diagonal self-energy matrix elements in the Kohn-Sham basis, namely $\langle \phi_n | \Sigma^{GW}_{XC}(\omega) | \phi_m \rangle$ ($n \ne m$) matrix elements, is ever calculated as in the quasiparticle self-consistent $\qsGW$ approach. Inspection of Eq.~\ref{eqn:qpequation} shows that the quasiparticle energies directly depend  on the quality of the diagonal approximation and on the shape of the input Kohn-Sham molecular orbitals. 

Partial self-consistency in such a diagonal approximation can be performed by reinjecting the calculated quasiparticle energies in the construction of $G$ and $W$, a popular scheme labeled $\evGW$. \cite{Shishkin2007,Blase2011,Kaplan2016} The $\evGW$ scheme was shown to reduce significantly the dependence on the starting input Kohn-Sham molecular orbitals, even though for some systems such a dependence remains strong. We will provide an example here below.

\subsection{ Quasiparticle energies as the poles of the $GW$ Green's function }

An alternative approach to finding the quasiparticle energies consists in considering directly the poles of the one-body Green's functions. 
Taking on general grounds the spectral representation of $G$ as given by Eq.~\ref{eqn:GreenFunc}, 
the expectation value of $G$ on a specific input molecular orbital $(\phi_n)$ reads:
\begin{align}
   \langle \phi_n |  G({\bf r},{\bf r}'; \omega ) | \phi_n \rangle =  \sum_{m}    \frac{  | \langle \phi_n | g_m \rangle |^2  }
   { {\omega} - \varepsilon_m  +i \eta \times \text{sgn}(\varepsilon_m  -E_F) }
\end{align}
Clearly, the poles of $\langle \phi_n | G | \phi_n \rangle$ are independent of the chosen one-body wavefunctions representation $\lbrace \phi_n \rbrace$. Further, concerning specifically the $GW$ Green's function: 
\begin{align}
    G^{-1}(\omega) = G^{-1}_{KS}(\omega)  + \Sigma_{XC}^{GW}(\omega)-V_{XC}^{DFT}
    \label{eqn:DysonG}
\end{align}
with $G_{KS}$ the input Kohn-Sham Green’s function, the search for the quasiparticle energies as the poles of $G$ does not require assuming the diagonality of the self-energy in the input $\lbrace \phi_n \rbrace$ basis. 
Namely, the   $\langle \phi_m | \Sigma_{XC}^{GW}(\omega) | \phi_n \rangle$ matrix is entirely considered and constructed with the fully dynamical self-energy as defined in eq.~\ref{eq:sigmaGW}. 

 The quasiparticle energies can then be calculated by extracting the dominant pole(s)  in the spectral representation of the diagonal matrix elements of the $GW$ Green's function in the available one-body molecular orbitals (MOs), typically the input Kohn-Sham MOs or the quasiparticle MOs of the previous iteration in a $\qsGW$ scheme: 
\begin{align}
    A_n^{GW}(\omega) =  \frac{1}{\pi} \left| \mathcal{I}m\left( \langle \phi_n | G(\omega) | \phi_n \rangle \right) \right|
    \label{eqn:spectral}
\end{align}
On formal grounds, this approach relies on the "quasiparticle" assumption, namely that $A_n^{GW}(\omega)$ will be dominated by a peak capturing most spectral weight, or a ``forest" of peaks that can be well described by a Lorentzian envelop providing the  $\varepsilon_n^{QP}$ quasiparticle energy, the $Z_n^{QP}$  spectral weight, and the associated   $\Gamma_n^{QP}$ lifetime: 
$$
    A_n^{fit}(\omega) =  \frac{1}{\pi} \left| \mathcal{I}m\left( \frac{Z_n^{QP}}{\omega -\left(\varepsilon_n^{QP} + i \Gamma_n^{QP}\right)}  \right) \right|
$$ 
In particular, the only constraint on the $\phi_n$ one-body orbital is that it should overlap significantly with the Lehman weight(s) $\lbrace g_m \rbrace$ associated with the n-th quasiparticle energy. As stated above however, the quasiparticle energy does not depend on the choice of $\phi_n$. 
Details about calculating and fitting the spectral function $A_n^{GW}(\omega)$ is described in Ref.~\citenum{Duchemin2020} and  in the Technical subsection below.

\begin{figure}[b]
	\includegraphics[width=8cm]{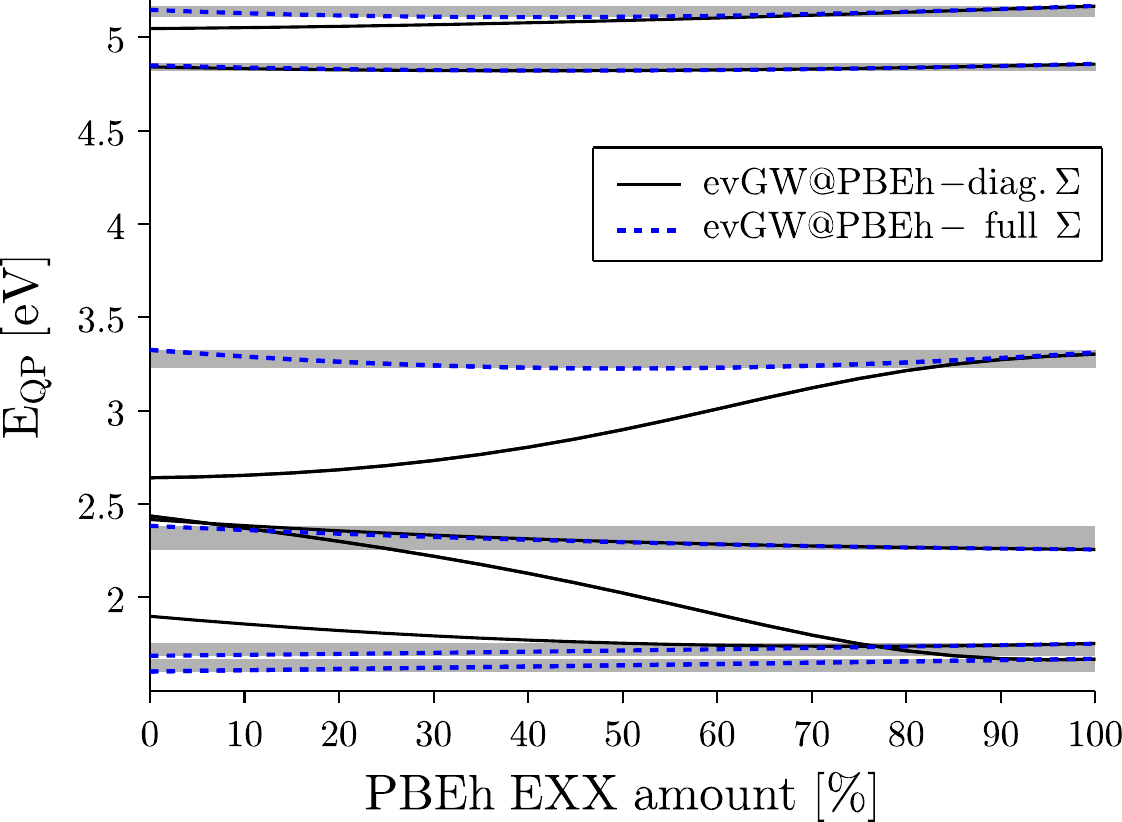}
	\caption{  Empty states energy levels for the  CO molecule as obtained from  $\evGW$ calculations with partial self-consistency on the eigenvalues. The energies are plotted as a function of the amount of exact exchange in the starting    Kohn-Sham    PBEh  functional used to generate the  input Kohn-Sham eigenstates. The traditional diagonal approximation on the self-energy operator (full black lines, $\evGW$-diag. $\Sigma$)  leads to a strong residual dependence on the starting functional for Rydberg states, while extracting the quasiparticle energies from the spectral function $A_n(\omega)$ (dashed blue lines; $\evGW$-full $\Sigma$) dramatically stabilizes the $\evGW$ results. Calculations performed at the aug-cc-pVTZ level. Grey areas indicate the small variations of the $\evGW$ energy levels extracted from $A_n^{GW}(\omega)$. We adopt here the CC3/aug-cc-pVTZ geometry of Ref.~\citenum{Sarkar2021}. }
	\label{fig:fig1}
\end{figure}

To illustrate the impact of releasing the diagonal approximation on the self-energy, and extracting the quasiparticle energies directly from the spectral function,  we  study the quasiparticle energies  associated with the low lying unoccupied energy levels of carbon monoxide (CO).  We keep here the input Kohn-Sham orbitals frozen, but perform a partial self-consistency on the quasiparticle energies, namely an $\evGW$ cheme. This allows   studying specifically the impact of the choice of a given set of input wavefunctions. The results are reported in Fig.~\ref{fig:fig1}. The  input Kohn-Sham orbitals are generated with the PBEh($\alpha$) global  hybrids, \cite{Perdew1996} with $\alpha$ ranging from zero (the PBE functional\cite{Perdew1996pbe}) to one (a hundred percent of exact exchange). Calculations are performed with the augmented aug-cc-pVTZ basis set \cite{Dunning89,Kendall92} to deal with diffuse states with positive energy. 

The standard $\evGW$ results, using eq.~\ref{eqn:qpequation}, are represented by full black lines ($\evGW$ diag-$\Sigma$; Fig.~\ref{fig:fig1}). Clearly, this standard scheme leads to a dramatic sensitivity of some of the lowest CO unoccupied energy levels with the amount of exact exchange in the starting Kohn-Sham functional, despite the self-consistency on the quasiparticle energies. 
In particular, the Kohn-Sham LUMO orbital becomes the LUMO+1 for a wide range of exact exchange percentage at the $\evGW$ diag-$\Sigma$ level, only to fall back to the LUMO level when starting with a PBEh hybrid functional with more that $\sim 75\%$ of exact exchange.
This is the signature that the associated Kohn-Sham wavefunctions strongly depend  on the DFT XC functional. Such a sensitivity was shown \cite{Villalobos2023} to originate from the diffuse nature of the unoccupied CO molecular orbitals that experience the significant change in the XC potential vacuum tail as a function of the amount of exact exchange.

We now  calculate the full self-energy matrix $\Sigma_{XC}^{GW}(\omega)$ in the Kohn-Sham basis to build the $GW$ Green's function (eq.~\ref{eqn:DysonG}), capturing the quasiparticle energies $\varepsilon_n^{QP}$ from the dominant peak in the related $A_n^{GW}(\omega)$.   
The results are represented by the dashed blue lines in Fig.~\ref{fig:fig1} ($\evGW$ full-$\Sigma$). Clearly, the obtained quasiparticle energies are much less sensitive to the choice of the input exchange-correlation (XC) functional and related one-body molecular orbitals (MOs).  This illustrates that even without self-consistency on the one-body MOs, extracting the quasiparticle energies from the pole(s) of the diagonal matrix elements of $G$, but considering the full self-energy matrix in the input Kohn-Sham basis when building $G$, allows   reducing dramatically the dependence of the quasiparticle energies on the input Kohn-Sham set of wavefunctions. 

We note that the standard $\evGW$ calculations, with the $\Sigma$-diagonal assumption, and  the approach bypassing the diagonality assumption, provide very similar results for input Kohn-Sham XC functionals containing above 80$\%$ of exact exchange. Such a value is close to the 75$\%$ found in an IP-tuning strategy for CO. \cite{Villalobos2023}   Benchmark calculations on molecular systems \cite{Rangel2016,McKeon2022} supported the conclusion that standard $GW$ calculations, with frozen input Kohn-Sham wavefunctions generated with an optimally-tuned Kohn-Sham functional, would lead to accurate $GW$ calculations and, subsequently, related Bethe-Salpeter optical spectra. \cite{McKeon2022,Kshirsagar2023} This is an indication that the approach relying on the $\langle \phi_n | G(\omega) | \phi_n \rangle$ spectral function, without assuming the diagonality of $\Sigma$ operator in the KS basis, yields results that are not only much less dependent on the input Kohn-Sham molecular orbitals, but also close to the best non-self-consistent calculations relying on   optimally tuned initial Kohn-Sham functionals.   

\subsection{ Quasiparticle self-consistent $GW$ approach with a Joint Approximate Diagonalization scheme }

While working with the spectral function, bypassing the diagonal self-energy approximation,  can reduce the impact of the input $\phi_n$ molecular orbitals (MO) on the quasipaticle energies, it is well documented that in many situations, self-consistency on the MOs becomes  required. This is e.g. the case of systems involving very localized \textit{d} or \textit{f} orbitals. Inaccurate KS MOs, of the kind obtained with purely local functionals, may lead to an erroneous density-matrix, spoiling the resulting Hartree and exchange potentials, and further to inaccurate response functions (susceptibility, screened Coulomb potential).  In addition,  subsequent post-processing of $GW$ results, such as calculating the optical spectrum within the Bethe-Salpeter equation formalism, can be very sensitive to the shape of the MOs. \cite{Kshirsagar2020,Kshirsagar2023}

We then now proceed to a full quasiparticle self-consistent scheme ($\qsGW$) that avoids relying on the diagonalization of an Hamiltonian  based on an optimal \textit{ansatz} static and symmetrized self-energy operator:
\begin{align}
  \Sigma^{\qsGW}_{nm} \simeq  \frac{1}{2}  \langle \phi_n |   \Sigma( \varepsilon_n ) + \Sigma( \varepsilon_m )   | \phi_m \rangle
  \label{eq:SigmaFaleev}
\end{align}
as introduced by Faleev and coworkers.\cite{Faleev2004,Schilfgaarde2006} 
 Following a common approach in signal processing, 
we adopt a \textit{ Joint Approximate Diagonalization (JAD) \cite{Cardoso1996} of the set of Green's function matrices taken at the quasiparticle energies}. Namely, we look for a unitary rotation $U$ within the input Kohn-Sham MOs that minimizes the off-diagonal matrix elements of the $\{G_n = G( \varepsilon_n^{QP} ) \} $ set of matrices. This joint minimization scheme can be formulated as:
\begin{equation}  
\argmin_{U}   {\mathcal F}(U,G_n)
\end{equation} 
with:
\begin{align}
     {\mathcal F}(U,G_n) &= \sum_n^{N_{orbs}} || \; \text{off-diag}(   U^{\dagger} G( \varepsilon_n^{QP} ) U) \; ||
\end{align}
where   $\lbrace U,G_n \rbrace$  are expressed in the Kohn-Sham MO basis.   The upper limit  $N_{\text{orbs}}$  indicates the number of Kohn-Sham orbitals that are allowed to mix, typically all occupied states and a large number of unoccupied states. As stated above,  working with the $GW$ Green's function allows   preserving the fully dynamical $GW$ self-energy as defined in eq.~\ref{eq:sigmaGW} without any approximation. 

Qualitatively, we thus look for the optimal one-body wavefunction basis that maximizes the diagonality of the one-body Green's functions taken at the quasiparticle energies.  While  the standard $\qsGW$ scheme relies on the definition of a static symmetrized self-energy, here the approximation lies in that there is in general no rotated $\lbrace U\phi_n^{KS} \rbrace$ basis that can strictly diagonalize all  $ G( \varepsilon_n^{QP} )  $ matrices at the same time. The two approximations are thus \textit{a priori} of different  nature, even though both aiming at constructing ``optimal" one-body quasiparticle $\lbrace \varepsilon_n^{QP}, \phi_n^{QP} \rbrace$ eigenstates.  

The self-consistency proceeds by using these one-body eigenstates to update the density matrix:  
\begin{align}
\gamma({\bf r},{\bf r}') = \sum_n \phi_n^{QP}({\bf r}) \phi_n^{QP}({\bf r}')^* \theta(E_F - \varepsilon_n^{QP})
\label{eq:gammaQP}
\end{align}
and associated charge-density, together with an updated  Green's function to start the new iteration:
    \begin{align}
    G^{QP}({\bf r},{\bf r}'; \omega ) =  \sum_{n}    \frac{   \phi_n^{QP}({\bf r})  \phi_n^{QP}( {\bf r}')^*  }{ {\omega} - \varepsilon_n^{QP} +i \eta \times \text{sgn}(\varepsilon_n -E_F) }  
 \label{eqn:GreenFuncQP0}
\end{align}
Likewise, an updated susceptibility is built replacing the KS eigenstates in eq.~\ref{eqn:xi0} by the quasiparticle eigenstates. This allows building an updated Hartree potential and  $\Sigma_{XC}^{GW}$ self-energy. 

\begin{figure}[t]
	\includegraphics[width=8.6cm]{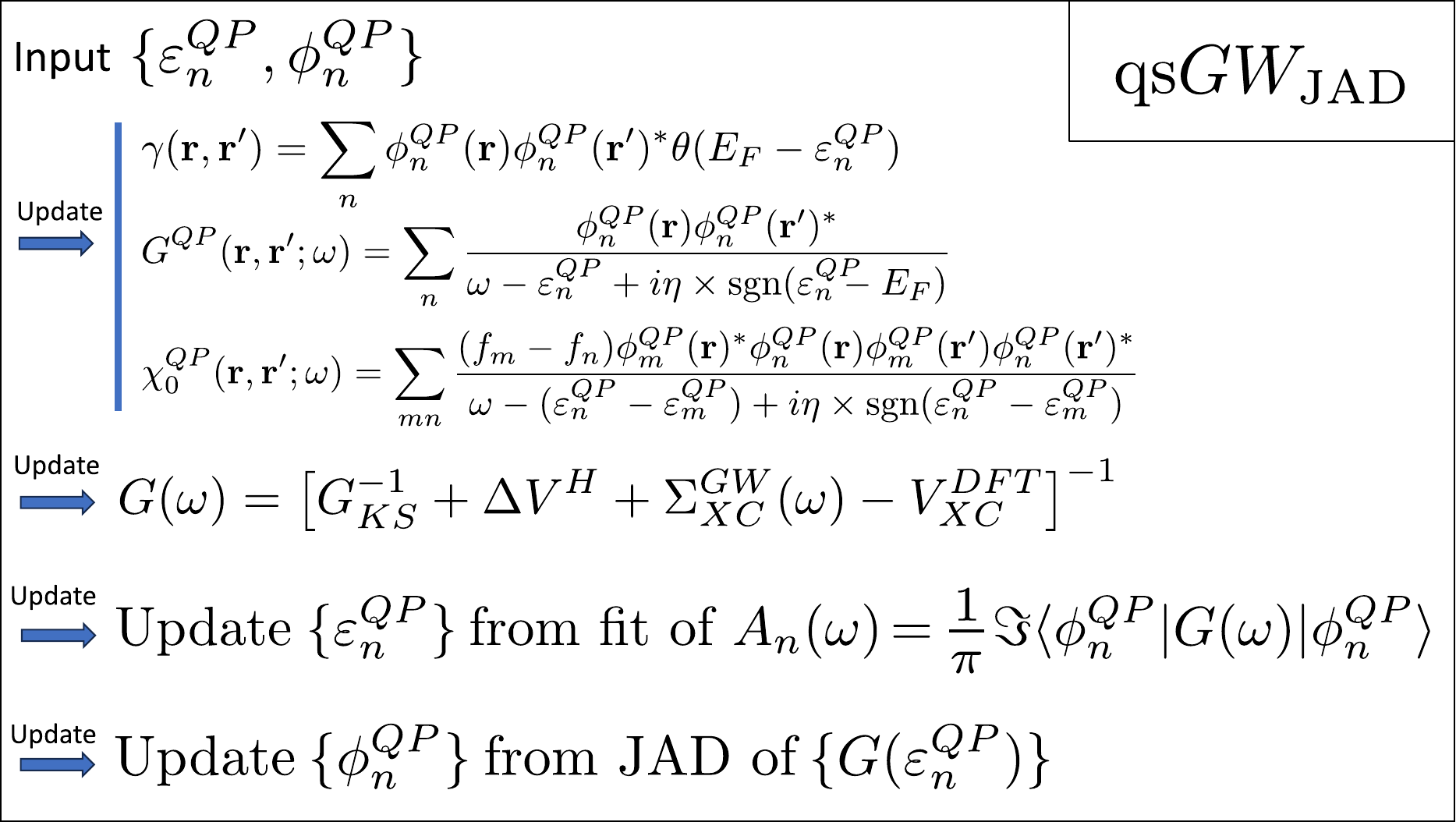}
	\caption{ Schematic representation of the $\qsGWJAD$ self-consistent scheme. }
	\label{fig:fig2}
\end{figure}

In the traditional $\qsGW$ scheme, $\Sigma_{XC}^{GW}$ would be reduced to the static symmetrized ansatz as given by  eq.~\ref{eq:SigmaFaleev} for direct diagonalization. In the present case, the full dynamical $\Sigma_{XC}^{GW}$ is used to build an updated $GW$ Green's function beyond the quasiparticle approximation: 
\begin{align}
    G^{-1}(\omega) = G^{-1}_{KS}(\omega)  + \Delta V_H + \Sigma_{XC}^{GW}(\omega)-V_{XC}^{DFT}
    \label{eqn:DysonG2}
\end{align}
where $\Delta V_H$ is the variation of the Hartree potential with respect to the input KS one. After extracting the new quasiparticle energies $\varepsilon_n^{QP}$ from the dominant pole(s) of the updated $G(\omega)$ and associated $A_n(\omega)$, the JAD of these new $G(\varepsilon^{QP}_n )$  allows  updating the quasiparticle one-body molecular orbitals to feed the next iteration. This scheme will be labeled here below the $\qsGWJAD$ scheme
(see schematic flow in Fig.~\ref{fig:fig2}). 
We emphasize that $G^{QP}$ is really written in the quasiparticle form, with a spectral function consisting of $N_{\text{basis}}$ $\delta$-functions with weight Z=1 at the available input quasiparticle energies, with $N_{\text{basis}}$ the size of the Kohn-Sham basis. $G^{QP}$ only serves to build the updated self-energy $\Sigma_{XC}^{GW}$. On the contrary, $G$ not only contains the quasiparticle peaks, with an  associated $Z \le 1$ spectral weight, but also the incoherent background. 

Since the JAD scheme uses the self-energy as it stands, it can be straightforwardly merged with any form of dynamical self-energy extending beyond the $GW$ approximation, including  renormalized singles, \cite{Weitao2019,Weitao2022,Golze2022} second-order screened exchange, \cite{Rinke2015} and various forms of vertex corrections \cite{Godby1994,Schindlmayr1998,Forster2024,Kresse2007,Romaniello2009,Leeuwen2014,Kresse2017,Berkelbach2019,Vlcek2019,Kutepov2021,Visscher2022,Forster2023,Patterson2024,Zgid2024,Ferretti2024} to the $GW$ self-energy. Likewise, the present approach can be merged with other forms of self-energy, such as the $GT$ approximation based on ladder diagrams, that has also been appraised for the calculation of the quasiparticle properties of  molecular systems. \cite{Yang2017,Loos2023}

\subsection{Modified  $\qsGW$ with a density-matrix beyond the quasiparticle approximation}

A simple alternative approach to the present $\qsGWJAD$ stems from the possibility to calculate the density-matrix $\gamma({\bf r}, {\bf r}')$  by integrating along the imaginary axis the time-ordered Green's function (eq.~\ref{eqn:DysonG2}) :
\begin{equation}
    \gamma({\bf r}, {\bf r}') =  \frac{1}{2} \left( \delta({\bf r}, {\bf r}') +
    \frac{1}{\pi}\int_{-\infty}^{+\infty} d\omega \; G({\bf r}, {\bf r}'; i\omega) \right)
    \label{eq:gamma2}
\end{equation}
and not from the rotated occupied $\lbrace \phi_n^{QP} = U \phi_n^{KS} \rbrace$ eigenstates. As such, the density-matrix in eq.~\ref{eq:gamma2} captures the contributions from the quasiparticle peaks and the incoherent background, as in a full $\scGW$ self-consistent scheme. This leads to an internal loop where starting from the Hartree and exchange operators built with the density matrix of eq.~\ref{eq:gammaQP}, the Hartree and exchange contributions to $G$ in eq.~\ref{eqn:DysonG2} are updated through eq.~\ref{eq:gamma2} keeping the correlation self-energy frozen. This does not cost significant time.  Such an approach will be labeled the $\gsGWJAD$ scheme, where the ${\gamma}s$ subscript indicates that the density-matrix is calculated as in the full self-consistent scheme. It remains that the $GW$ self-energy is constructed with $G^{QP}$ and the non-interacting susceptibility $\chi_0$ built from one-body molecular orbitals. This approach adopts thus features from both the $\qsGW$ and the fully-self-consistent $\scGW$ schemes.

\subsection{ Technical details }

Our calculations are performed with the {\sc{beDeft}} (beyond-DFT) package \cite{Duchemin2020,Duchemin2021} implementing the $GW$ and Bethe-Salpeter equation (BSE) formalisms with Gaussian basis sets and  Coulomb-fitting (RI-V) resolution-of-the-identity techniques. \cite{Whitten1973,Vahtras1993,Ren2012,Duchemin2017}  We exploit in particular a recently improved  analytic continuation (AC) scheme combined with the contour-deformation approach  that allows  calculating accurately the $GW$ self-energy, even for levels located far away from the gap.  \cite{Duchemin2020}  The independent electron  susceptibility $\chi_0(z)$  and related RPA screened Coulomb potential $W(z)$ are calculated for an optimized grid of $N_{\omega}$ frequencies ($z=i\omega$) along the imaginary frequency axis ($N_{\omega} = 14)$, completed by a coarse grid of complex frequencies ($z=\omega+i\eta$) above the real-axis.  This set of calculations allows setting an accurate analytic continued-fraction expression for the screened Coulomb potential, and consequently an analytic form for the self-energy and associated Green's function. The stability of the quasiparticle energies with respect to the sampling grid, from core to unoccupied levels, was extensively studied in Ref.~\citenum{Duchemin2020}. Such an improved AC scheme was recently exploited by several groups in the study of core levels. \cite{Duchemin2020,Kehry2023,Barrueta2023}  
Besides all occupied states, off-diagonal self-energy matrix elements are constructed with unoccupied states within an energy window of 200 eV above the gap. The JAD minimization process with respect to the unitary $U$ operator matrix elements does not represent by itself a significant computational effort.

To allow comparison with previous $\qsGW$ and $\scGW$ studies of the highest-occupied molecular orbital (HOMO) energies for the so-called $GW$100 molecular test set, \cite{Caruso2016}  with   associated reference CCSD(T) data, \cite{Krause2015} we adopt the def2-TZVPP basis set \cite{Weigend2005}  and the corresponding optimized auxiliary basis set. \cite{Weigend1998}   Input Kohn-Sham eigenstates are generated by the Orca package. \cite{Neese2020} We adopt the PBE0 functional \cite{Perdew1996,Adamo1999} as the mean-field Kohn-Sham starting point. Following the $\qsGW$ and $\scGW$ reference calculations from Ref.~\citenum{Caruso2016}, we exclude molecules containing fifthrow atoms for which all-electron def2-TZVPP basis sets are not available, reducing the GW100 test set to 93 molecules [see Table S1 in the Supplementary Material (SM)].

\section{Results }

\begin{figure}[t]
\includegraphics[width=8cm]{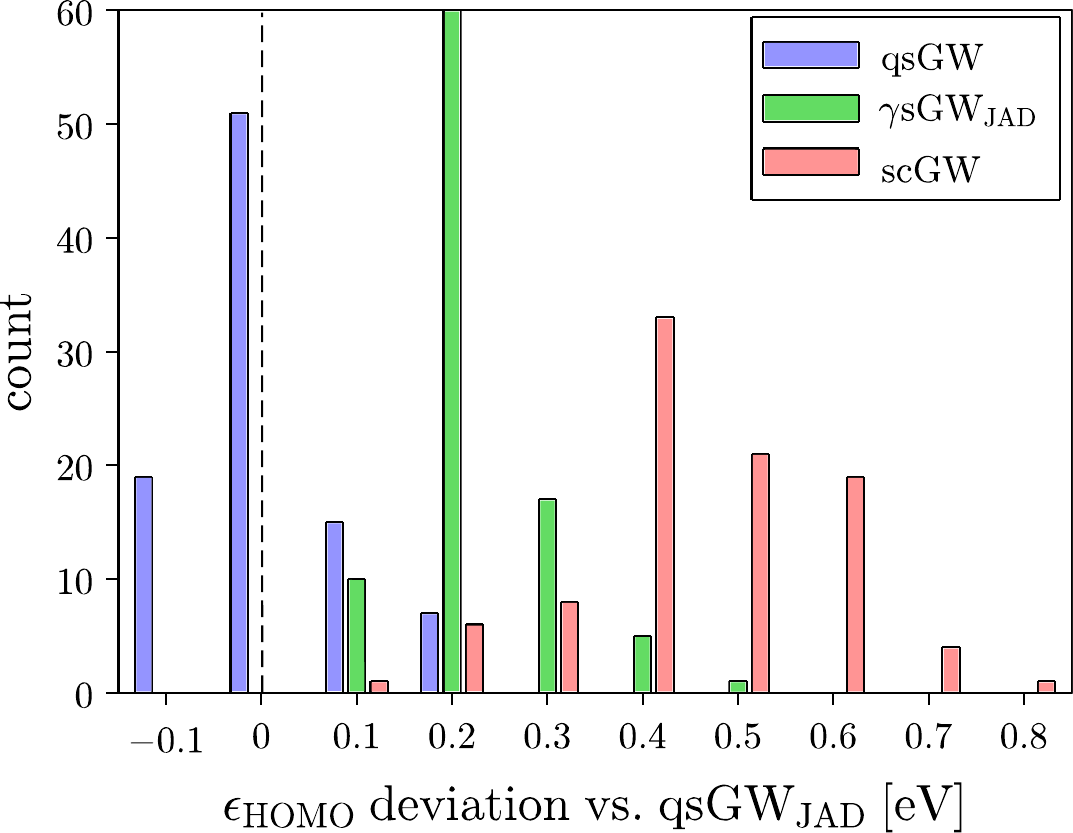}
	\caption{ Bar plot of the $\varepsilon_{HOMO}^{\qsGW}$ (blue), $\varepsilon_{HOMO}^{\gsGWJAD}$ (green), and $\varepsilon_{HOMO}^{\scGW}$ (red) energies, taking as a reference the present $\varepsilon_{HOMO}^{\qsGWJAD}$ values. The $\qsGW$ and $\scGW$ data are from Ref.~\citenum{Caruso2016}, considering the full set of 93 molecules reported, except  the $C_8H_8$ cyclooctatetraene molecule which really stands as an outlier taking  the $\qsGW$ value of Ref.~\citenum{Caruso2016} (see Note~\citenum{C8H8note}). The $\qsGWJAD$ and $\gsGWJAD$ data are available in the Table S1 of the SM.  }
	\label{fig:fig3}
\end{figure}

We start by providing in Fig.~\ref{fig:fig3} (blue) a bar plot of the difference $(\varepsilon_{HOMO}^{\qsGW} - \varepsilon_{HOMO}^{\qsGWJAD})$ for the set of 93 molecules extracted from the $GW$100 test set.  The $\qsGWJAD$ data are available in the Table S1 of the Supplementary Material (SM) while the $\varepsilon_{HOMO}^{\qsGW}$ are taken from Ref.~\citenum{Caruso2016}. The present $\qsGWJAD$ scheme yields as expected results slightly different from the original $\qsGW$ scheme. However, the associated mean-absolute-error (MAE) and mean-signed-error (MSE)  are small, amounting to 65 meV and 3 meV, respectively. These deviations   change to 56 meV and 13 meV if one removes the $C_8H_8$ outlier\cite{C8H8note} for which the $\qsGW$ HOMO energy seems really off in the calculations of Ref.~\citenum{Caruso2016}. For sake of comparison, we  also plot the difference $(\varepsilon_{HOMO}^{\scGW} - \varepsilon_{HOMO}^{\qsGWJAD})$ (red) on the basis of the $\scGW$ data from Ref.~\citenum{Caruso2016}. Such a comparison shows that the two $\qsGW$ schemes are  very close to each other as compared to the differences between the $\qsGW$ and $\scGW$ data. The MAE and MSE between $\qsGWJAD$ and $\scGW$ are indeed much larger, amounting both to 0.45~eV. Such an agreement between $\qsGW$ and $\qsGWJAD$ data is rather remarkable given the very different nature of the approximations on which the two schemes are hinging. 

\begin{figure}[t]
	\includegraphics[width=8cm]{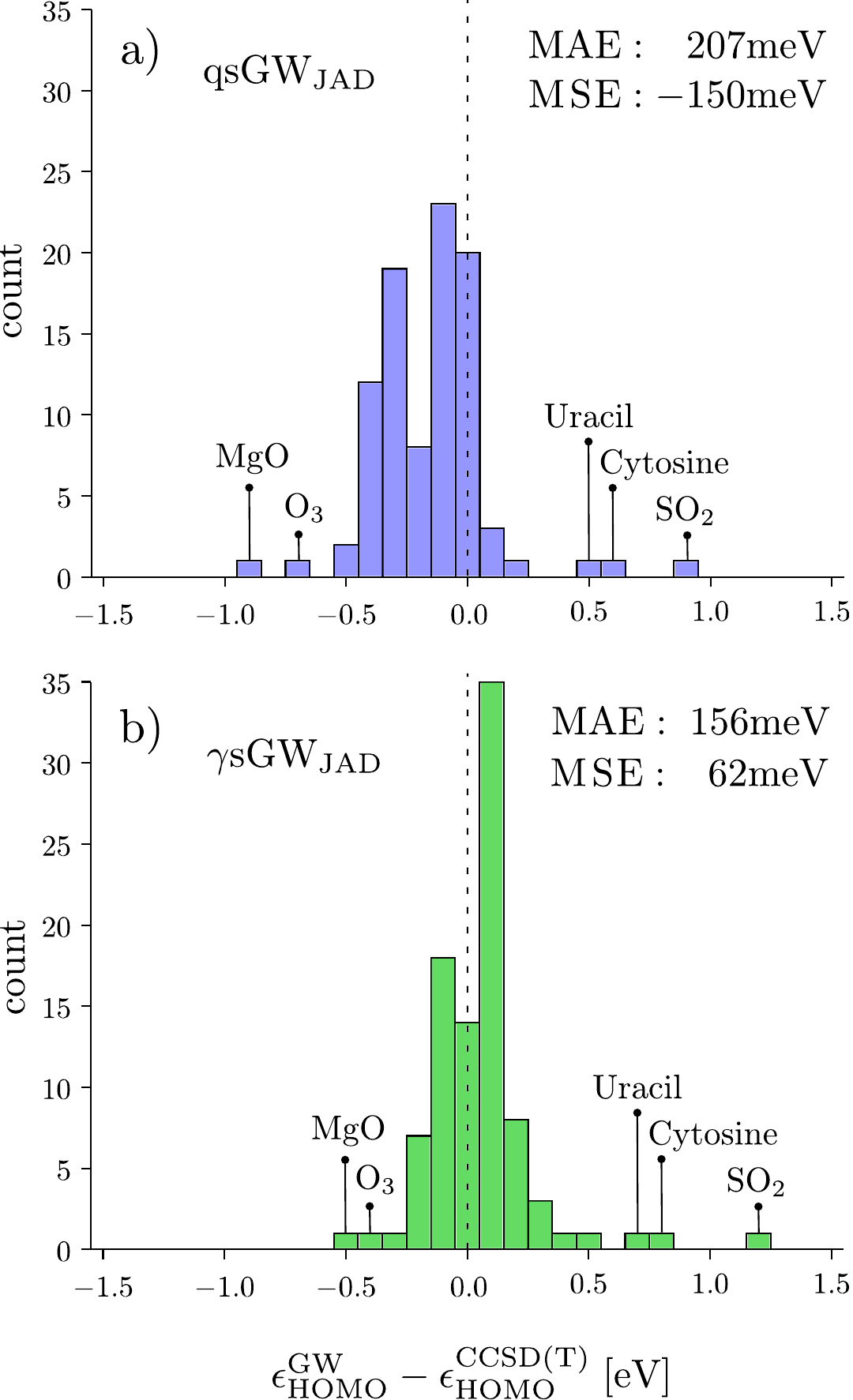}
	\caption{ Histogram of the $(\varepsilon_{HOMO}^{GW} - \varepsilon_{HOMO}^{\text{CCSD(T)}})$ difference for (a) the $\qsGWJAD$ scheme, and (b) the $\gsGWJAD$ for 93 molecules out of the $GW$100 test set (see text). Associated mean-absolute-errors (MAE) and mean-signed-errors (MSE)  are indicated in each case. }
	\label{fig:fig4}
\end{figure}

We now compare in Fig.~\ref{fig:fig4}(a) the present $\qsGWJAD$ data to the CCSD(T) reference. The $\qsGWJAD$ scheme  yields too deep HOMO levels, that is too large ionization potentials (IP),  with MSE/MAE of -0.15 eV/0.21 eV, respectively. As expected, this is nearly identical to the MSE/MAE of  -0.15 eV/0.22 eV   characterizing the difference between $\qsGW$ and CCSD(T). \cite{SRGqsGWnote} In particular, the error associated with the outliers  (see corresponding names  in Fig.~\ref{fig:fig4}), are  close to what was found for the standard $\qsGW$ scheme (see values in Table S1 in the SM), except for the $C_8H_8$ outlier as discussed above. 

We further examine in Fig.~\ref{fig:fig4}(b) the present $\gsGWJAD$ data as compared to the CCSD(T) reference. As explained above, the update of the Hartree and exchange potentials are here performed by calculating the density matrix through an integral along the imaginary axis of the full Green's function $G=[ G^{-1}_{KS}  + \Delta V_H + \Sigma_{XC}^{GW}-V_{XC}^{DFT}]^{-1}$ (eqn.~\ref{eqn:DysonG2}), and not as a sum over the occupied  $\lbrace \phi_n^{QP} = U \phi_n^{KS} \rbrace$ rotated one-body wavefunctions. The associated results are found to be in better agreement with the CCSD(T) data. The corresponding MAE and MSE reduce to 156 meV and 62 meV, respectively. While $\qsGW$ and $\scGW$ show a deviation with respect to the CCSD(T) HOMO energy by -0.15 eV and 0.30 eV\cite{Zgidnote} (MSE values) respectively, this intermediate scheme yields data in between the two self-consistent $GW$ approaches, closer to CCSD(T) reference. This is confirmed by Fig.~\ref{fig:fig3} showing in green that the $\gsGWJAD$ scheme yields results intermediate between that of  $\qsGW$ and $\scGW$.  The outliers remain the same as the one identified within the $\qsGW$ and $\qsGWJAD$ schemes. 

\begin{figure}[t]
	\includegraphics[width=8cm]{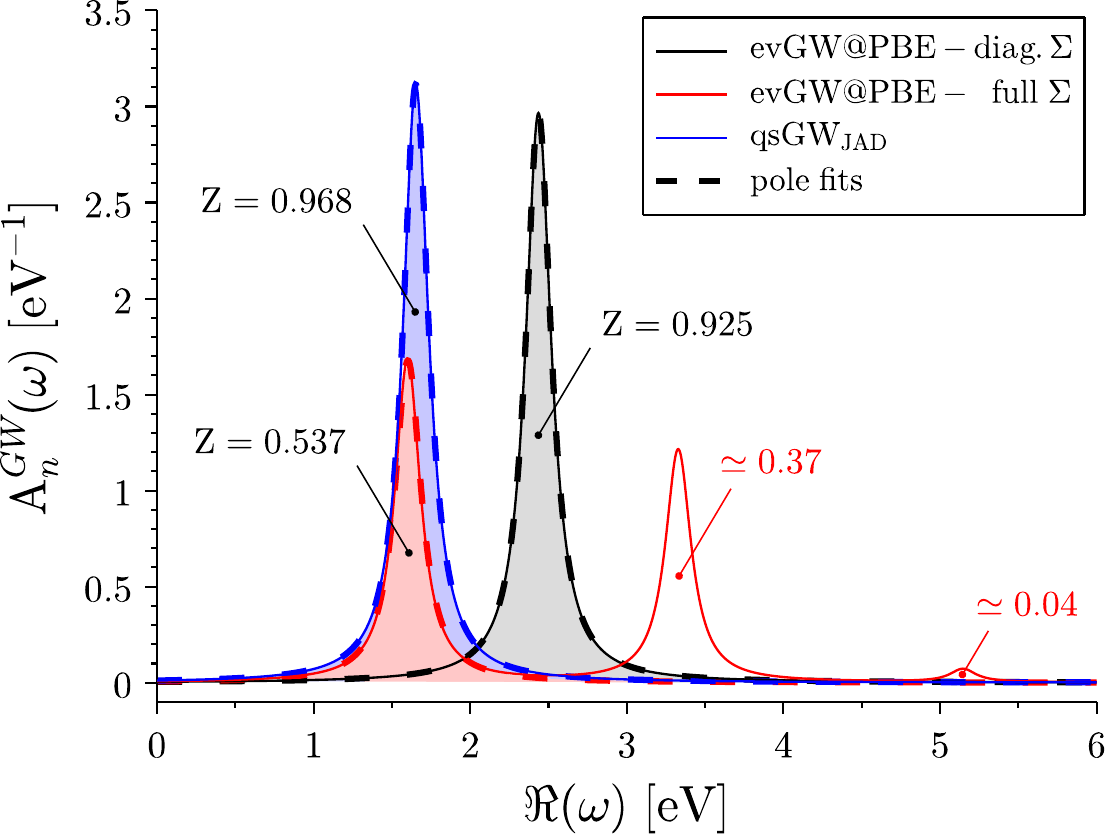}
	\caption{ Plot of the spectral function $A_n^{GW}(\omega)$ associated with the carbon-monoxide  LUMO level.
 We compare the (grey) ev$GW$@PBE-diag.$~\Sigma$, (red) ev$GW$@PBE-full~$\Sigma$ and (blue) $\qsGWJAD$ schemes. Dashed lines are the lorentzian fits of the principal poles with their associated $Z$ spectral weights. An estimation of the spectral weight associated to secondary poles in the ev$GW$@PBE-full~$\Sigma$ case is also provided.}
	\label{fig:fig5}
\end{figure} 

 We close our discussion of the JAD process by plotting in Fig.~\ref{fig:fig5} the spectral function $A_n^{GW}(\omega)$ associated with the carbon-monoxide LUMO at the PBE Kohn-Sham level. We compare the $A_n^{GW}(\omega)$ obtained with   (a) the ev$GW$@PBE-diag.~$\Sigma$ scheme (in grey), (b)    
the ev$GW$@PBE-full~$\Sigma$ scheme (in red), and (c) the $\qsGWJAD$ process (in blue).  While the ev$GW$-diag-$\Sigma$ scheme yields a single quasiparticle peak, but located $\sim$0.8 eV away from the $\qsGWJAD$ pole, the $\evGW$-full~$\Sigma$  scheme yields several structures, the strongest one ($Z \simeq 0.54$) defining the quasiparticle energy. The comparison between these two spectral functions indicates that the self-energy operator is strongly non-diagonal in the input PBE Kohn-Sham basis. If we now rotate the input molecular orbitals using the JAD process, the final spectral function becomes dominated by a single quasiparticle peak. Contrary to the $\evGW$-diag.~$\Sigma$ scheme, this is not obtained by enforcing artificially the diagonality of the self-energy operator in the one-body MO basis, but by finding the rotated MOs that maximalize the diagonality of the $G(\varepsilon_n^{QP})$ family of matrices, without any approximation on the self-energy operator. 

We note that the $\evGW$-full~$\Sigma$ scheme, that does not update the one-body MOs but build $G$ from the full self-energy in the Kohn-Sham basis, yields a main quasiparticle energy in much better agreement with   $\qsGWJAD$   as compared to the $\evGW$-diag.~$\Sigma$ result. However, the presence of two poles with significant weights may give rise to instabilities, with a dominant spectral weight jumping from one pole to another, with discontinuity of the quasiparticle energy, in a self-consistent process or upon varying a structural parameter (e.g. bond length in the study of a dissociation curve). \cite{Veril2018,Berger2021,Monino2022,Marie_2023}

\section{Conclusions}

We have proposed an alternative approach to quasiparticle self-consistent $GW$ calculations relying on the Joint Approximate  Diagonalization  (JAD) of the $GW$ Green's function $G(\varepsilon_n^{QP})$ taken at the available $\varepsilon_n^{QP}$ quasiparticle energies (input Kohn-Sham energies or previous iteration values in a self-consistent scheme). Such an approach does not rely on the set-up of a symmetrized and static \textit{ansatz} self-energy operator. In the JAD scheme, the approximation lies in the fact that there is no rotated one-body set of molecular orbitals that can strictly diagonalize all $G(\varepsilon_n^{QP})$ matrices, with $n$ ranging from core  to unoccupied levels in a very large energy range (a few hundred eVs). Remarkably, even though relying on approximations of seemingly very different nature, this alternative quasiparticle self-consistent scheme provides quasiparticle energies close to the one obtained with the standard $\qsGW$ scheme. We cannot exclude potential relations between the principles behind the construction of the effective static symmetrized self-energy of the standard $\qsGW$ scheme, and the present approximate diagonalization scheme of the $G(\varepsilon_n^{QP})$ matrices.  

We further tested a self-consistent scheme that extends beyond the quasiparticle framework by constructing the  density matrix from the $GW$ Green's function through an integral along the imaginary axis, capturing not only the spectral weight of the quasiparticles but also that of the background. Such an approach yields results located in between $\qsGW$ and $\scGW$ data, in better agreement with reference CCSD(T) calculations for the set of systems considered. Such a variant can be merged with the standard $\qsGW$ scheme but requires calculating the self-energy operator $\Sigma_{XC}^{GW}(i\omega)$ for a set of frequencies along the imaginary axis according to the chosen quadrature. 

 \section*{SUPPLEMENTARY MATERIAL}
See the Supplementary Material for a complete Table of the $GW$100 HOMO energies at the various self-consistent levels. 

\begin{acknowledgments}
The authors are indebted to insightful discussions with Pierre-Fran\c{c}ois Loos and Antoine Marie. The project received support from the French Agence Nationale de la Recherche (ANR) under contract ANR-20-CE29-0005 (BSE-Forces). 
Computational resources generously provided by  the national HPC facilities under contract GENCI-TGCC A0110910016 are acknowledged.
\end{acknowledgments}

\section*{Data Availability Statement}

The data that support the findings of this study are available within the article  and its supplementary material.

\vskip 2cm
\bibliography{xavbib.bib}

\end{document}